\documentclass[12pt,eqsecnum]{article}
\usepackage[dvips]{graphicx}
\usepackage{amssymb}
\usepackage{amsmath}
\usepackage{epstopdf}
\makeatletter

\@addtoreset{equation}{section}
\makeatletter

\newcommand{\D}{{\rm d}}

\newcommand{\dalm}{\kern1pt\vbox{\hrule height 0.9pt\hbox{\vrule width
0.9pt\hskip 2.5pt\vbox{\vskip 5.5pt}\hskip 3pt\vrule width 0.3pt}\hrule height
0.3pt}\kern1pt}

\def\b2hat{ {\hat b}_2 }

\begin{document}

\begin{titlepage}
\vfill
\begin{flushright}
\today
\end{flushright}

\vfill
\begin{center}
\baselineskip=16pt
{\Large\bf 
Gauss-Bonnet braneworld redux: \\
A novel scenario for the bouncing universe
}
\vskip 0.5cm
{\large {\sl }}
\vskip 10.mm
{\bf Hideki Maeda} \\

\vskip 1cm
{
	Centro de Estudios Cient\'{\i}ficos (CECs), Casilla 1469, Valdivia, Chile \\
	\texttt{hideki@cecs.cl}

     }
\vspace{6pt}
\end{center}
\vskip 0.2in
\par
\begin{center}
{\bf Abstract}
 \end{center}
\begin{quote}
We propose a new scenario for the bouncing universe in a simple five-dimensional braneworld model in the framework of Einstein-Gauss-Bonnet gravity, which works even with ordinary matter on the brane.
In this scenario, the so-called branch singularity located at a finite physical radius in the bulk spacetime plays an essential role.
We show that a three-brane moving in the bulk may reach and pass through it in spite of the fact that it is a curvature singularity.
The bulk spacetime is extended beyond the branch singularity in the $C^0$ sense and then the branch singularity is identified as a massive thin shell.
From the bulk point of view, this process is the collision of the three-brane with the shell of branch singularity.
From the point of view on the brane, this process is a sudden transition from the collapsing phase to the expanding phase of the universe.
The present result opens a completely new possibility to achieve the bouncing brane universe as a higher-curvature effect.
  \vfill
\vskip 2.mm
\end{quote}
\end{titlepage}




\tableofcontents

\newpage 

\section{Introduction}
\label{sec1}

Initial singularity is one of the most serious problems in modern cosmology.
It is inevitable under several physically reasonable conditions in general relativity as a consequence of the singularity theorems~\cite{singularity,singularity2}.
Since the quantum effect of gravity dominates close to curvature singularities, the completion of the quantum theory of gravity is necessary to answer to this problem.
Although the full quantum theory of gravity is not available at present, (Super)string/M-theory is one of the most promising candidates. 
Given the present circumstances, the results in the low-energy effective theory must be useful to gain insights and suggestions for the problem. 
As a possibility to avoid the initial singularity, the bouncing universe is one intriguing scenario~\cite{gv1993}.
Because the effective energy-momentum tensor given from the semiclassical correction may violate the energy conditions, the quantum-gravity effect could cause the big bounce of the universe, a transition from the contracting phase to the expanding phase.
(See~\cite{bounce} for a review.)

Higher-dimensional cosmology based on string-generated gravity models is an ambitious way  to explain problems in modern cosmology such as dark energy or initial singularity in a harmonic way by the effect of extra dimensions. 
After the publication of the papers by Randall and Sundrum~\cite{Randall}, string-inspired braneworld cosmology, in which the observable universe is a (3+1)-dimensional timelike hypersurface (three-brane) embedded in a higher-dimensional anti-de~Sitter (AdS) bulk space-time, has been investigated enthusiastically. (See~\cite{braneworld} for a review.) 
The dynamics of the Friedmann-Robertson-Walker (FRW) brane universe has been fully investigated to explain cosmic inflation in the early universe or dark energy.
In the present paper, we focus on the possibility of the bouncing brane universe.

Among two Randall-Sundrum models, we consider the single-brane model.
Assuming a perfect fluid obeying an equation of state $p=(\gamma-1)\rho~(0<\gamma\le 2)$ on the FRW brane in the simplest braneworld with $Z_2$ symmetry~\cite{krausida}, we can show that the bouncing universe is not realized for $2/3<\gamma \le 2$.
This implies that, although the condition is milder than the standard four-dimensional FRW cosmology, a matter field with negative pressure violating the strong energy condition is still necessary for the big-bounce.
Moreover, the bouncing universe requires the negative mass parameter in the Schwarzschild-Tangherlini-type bulk solution which causes a naked singularity. 

Nevertheless, there are models providing the bouncing universe on the brane with ordinary matter.
One possible way is to introduce a matter field in the bulk spacetime.
The low-energy effective theory of (Super)string/M-theory predicts several fields in the bulk spacetime such as a dilaton field or gauge fields.
For example, by the dimensional reduction from the 11-dimensional Ho{\v r}ava-Witten model to the five-dimensional bulk spacetime, a dilaton field and a U(1) gauge field appear~\cite{hw1996}.
In the presence of a U(1) gauge field in the bulk, the bouncing universe is generically realized on the brane~\cite{bounce-U(1)}.
However, it is expected that the bulk spacetime suffers from the mass inflation instability~\cite{pi1990,massinflation} when the brane approaches the inner Cauchy horizon in the bulk~\cite{hm2003}.
In this research direction, a model of the bouncing brane universe without a bulk inner horizon was constructed in the presence of a SU(2) Yang-Mills field, in which the bulk spacetime is regular and free-from singularities~\cite{om2003,om2004}.

In the present paper, we reconsider the possibility of the bouncing brane universe with ordinary matter on the brane.
As mentioned above, this is impossible in the standard Randall-Sundrum scenario based on general relativity unless the extra dimension is spacelike.
Actually the bouncing universe is generically realized if the extra dimension is timelike~\cite{ss2003}, but we do not consider this radical possibility here.
Instead, we consider the braneworld in the framework of Einstein-Gauss-Bonnet gravity.
While the bouncing universe in the Gauss-Bonnet braneworld was discussed also in~\cite{mp2010} with a U(1) gauge field in the bulk, we consider the vacuum bulk here.

Einstein-Gauss-Bonnet gravity is a natural higher-dimensional generalization of general relativity as a quasilinear second-order theory.
Also, the quadratic Lanczos (Gauss-Bonnet) Lagrangian appears in the low-energy limit of heterotic string theory together with a dilaton~\cite{Gross}.
The early stage of the research history of the Gauss-Bonnet braneworld was somehow winding.
There was a disagreement among the authors on the junction condition for the bulk spacetime and hence the resulting Friedmann equation on the brane.
This confusion was settled finally by the establishment of the generalized Isreal junction condition by Davis~\cite{davis2003} and independently by Gravanis and Willison~\cite{GB-junction}. 
(See also~\cite{onGB-juncton} and the comment in~\cite{davis2003}.)
The correct Friedmann equation in the Gauss-Bonnet braneworld was first derived by Charmousis and Dufaux in~\cite{GBbrane-FRWequation} and the dynamics of the Friedmann brane has been investigated by many authors up to now~\cite{GBbrane-dynamics,brownPhD}. 
(See~\cite{GBbrane-equations} for the covariant gravitational equations on the brane.) 
Also, its viability~\cite{GBbrane-viability}, creation of the universe~\cite{GBbrane-creation}, cosmological perturbations~\cite{GBbrane-perturbation} have been investigated.
The maximally symmetric brane was studied in~\cite{GBbrane-(A)dSbrane}.
Other aspects of the Gauss-Bonnet braneworld have also been studied~\cite{GBbrane-stringy}.
(See Sec. 5.7 in~\cite{Review} for a review.)
The purpose of the present paper is to point out a new and intriguing dynamical property of the Gauss-Bonnet brane universe.

In Einstein-Gauss-Bonnet gravity, there is an exact  symmetric vacuum solution which is a generalization of the Schwarzschild-Tangherlini solution in general relativity~\cite{bd1985,GBBH}.
One of the most characteristic properties of this solution is the existence of a curvature singularity at a nonzero physical radius for negative mass parameter, which is called the {\it branch singularity}.
The qualitative behavior of the Gauss-Bonnet brane universe has been analyzed by many authors; however, the role of the branch singularity in this context has not been clarified yet.
In the general relativistic case, the moment when the brane reaches the central singularity in the bulk corresponds to the big-bang (or big-crunch) time on the brane.
Therefore, one might naively think that the brane universe ends in (or starts from) some curvature singularity on the brane when it hits the branch singularity in the bulk.
However, we will show that it is not the case.

In the present paper, we will show that the branch singularity is harmless in the sense that a finite body is able to reach there without being crushed to a point or ripped apart.
We will also show that the vacuum bulk spacetime can be extended beyond the branch singularity in the  $C^0$ sense and then the branch singularity is considered as a massive thin shell.
As a consequence, it is concluded that the brane may reach and pass through the branch singularity into the extended region of the spacetime.
This process provides a new scenario for the bouncing universe on the brane.

The rest of the present paper is constituted as follows.
In the following section, we present the bulk spacetime and the dynamical equation for the FRW brane in Einstein-Gauss-Bonnet gravity.
In Sec.~III, we explain our new scenario for the bouncing universe.
In Sec.~IV, the weakness of the branch singularity is shown.
Concluding remarks and discussions including future prospects are summarized in Sec.~V.
In Appendix A, we present the geometric quantities of the bulk spacetime.
In Appendix B, the asymptotic behavior of the brane universe for $a\to \infty$ is presented.

Our basic notation follows~\cite{wald}.
The convention for the Riemann curvature tensor is $[\nabla _\rho ,\nabla_\sigma]V^\mu ={R^\mu }_{\nu\rho\sigma}V^\nu$ and $R_{\mu \nu }={R^\rho }_{\mu \rho \nu }$.
The Minkowski metric is taken as diag$(-,+,+,+,+)$, and Greek indices run over all spacetime indices.
We adopt the units in which only the five-dimensional gravitational constant $G_5$ is retained.

\section{Preliminaries}
We consider the following five-dimensional action for the bulk spacetime:
\begin{equation}
I=\frac{1}{2\kappa_5^2}\int d^5x\sqrt{-g}\biggl(R-2\Lambda+\alpha{L}_{\rm GB}\biggl), \label{action}
\end{equation}
where $\Lambda$ is the cosmological constant and $\kappa_5$ is defined by the five-dimensional gravitational constant  $G_5$ as $\kappa_5:=\sqrt{8\pi G_5}$. 
The Gauss-Bonnet term ${L}_{\rm GB}$ is a combination of the Ricci scalar $R$, the Ricci tensor $R_{\mu\nu}$ and the Riemann
tensor $R^\mu{}_{\nu\rho\sigma}$ as
\begin{equation}
{L}_{\rm GB}:=R^2-4R_{\mu\nu}R^{\mu\nu}+R_{\mu\nu\rho\sigma}R^{\mu\nu\rho\sigma},
\end{equation}
which does not give any higher-derivative (more than the second-derivative) term in the field equations.
The constant $\alpha$ in the action is the coupling constant of the Gauss-Bonnet term and for $\alpha\to 0$ our model reduces to the Randall-Sundrum model. 

The gravitational equation given from the action (\ref{action}) is
\begin{equation}
{G}^\mu{}_\nu +\alpha {H}^\mu{}_\nu +\Lambda \delta^\mu{}_\nu= 0, \label{beq}
\end{equation}
where
\begin{align}
{G}_{\mu\nu}:=&R_{\mu\nu}-{1\over 2}g_{\mu\nu}R, \\
{H}_{\mu\nu}:=&2\Bigl(RR_{\mu\nu}-2R_{\mu\alpha} R^\alpha{}_\nu -2R^{\alpha\beta} R_{\mu\alpha\nu\beta} + R_\mu{}^{\alpha\beta\gamma}R_{\nu\alpha\beta\gamma} \Bigr) - {1\over 2} g_{\mu\nu} {L}_{\rm GB}.
\end{align}
The Gauss-Bonnet term in the action is obtained in the low-energy limit of heterotic superstring theory together with a dilaton in ten dimensions~\cite{Gross}, in which case $\alpha$ is regarded as the inverse string tension and positive definite. 
We therefore assume $\alpha > 0$ throughout this paper.
We also assume $\Lambda<0$ and $1+4\alpha\Lambda/3> 0$ in addition, the latter of which ensures the existence of nondegenerate maximally symmetric vacuum solutions.

\subsection{Bulk solution}
\label{bulk}
In this system, a vacuum solution is obtained as a warped product manifold ${\cal M}^5 \approx M^2\times K^{3}$, where $K^{3}$ is a three-dimensional space of constant curvature. 
In the equations which follow, $k$ denotes the curvature of $K^{3}$ and takes the values $1$ (positive curvature), $0$ (zero
curvature), and $-1$ (negative curvature). 
The metric of the vacuum solution is given by 
\begin{align}
ds_5^2=&g_{\mu\nu}dx^\mu dx^\nu \nonumber \\
=& -h(r)dt^2+\frac{dr^2}{h(r)}+r^2 \left[d\chi^2+f_k(\chi)^2(d\theta^2+\sin^2\theta d\phi^2) \right],\label{h-eq} \\
h(r) :=& k+\frac{r^2}{4\alpha}\Biggl(1\mp\sqrt{1+\frac{\alpha \mu}{r^{4}}+\frac43\alpha\Lambda}\Biggr),\label{horizon-2}
\end{align}
where $\mu$ is constant, $f_0(\chi):=\chi$, $f_1(\chi):=\sin\chi$,  and $f_{-1}(\chi):=\sinh\chi$~\cite{bd1985,GBBH}.
The relation between $\mu$ and the global mass parameter $M$ is 
\begin{align}
\mu=\frac{16\kappa _5^2M}{3V_{3}^{(k)}},
\end{align}
where the constant $V_{3}^{(k)}$ is a unit volume of $K^{3}$ if it is compact.
In the asymptotically flat case ($k=1$), $M$ gives the ADM mass.
 
It is seen that the solution has two branches corresponding to the sign in the metric function  $h(r)$.
We call the family with the minus (plus) sign the GR branch (non-GR branch).
Only the GR branch solution has the general relativistic limit as
\begin{equation}
\lim_{\alpha\to 0}h(r)=k-\frac{\mu}{8r^2}-\frac16 \Lambda r^2.
\end{equation}
The maximally symmetric vacuum in the non-GR branch was shown to be unstable~\cite{bd1985,cp2008} and so we only consider the solution in the GR branch in the present paper.

The global structures of this spacetime depending on the parameters have been clarified~\cite{tm2005}.
In this spacetime, there are two classes of curvature singularity for $\mu \ne 0$.  
One is the central singularity at $r=0$ and the other is the branch singularity at $r=r_{\rm b}(>0)$, where the term inside the square-root in the metric function~(\ref{horizon-2}) vanishes. 
$r_{\rm b}$ is explicitly given by
\begin{equation}
r_{\rm b}:=\biggl(-\frac{3\alpha\mu}{3+4\alpha\Lambda}\biggl)^{1/4}.
\end{equation}
The branch singularity exists if $\mu$ is negative.
The metric and its inverse are finite at $r=r_{\rm b}$ (but their derivatives blow up) and the metric becomes complex and hence unphysical at $r<r_{\rm b}$.

\begin{figure}[htbp]
\begin{center}
\includegraphics[width=0.6\linewidth]{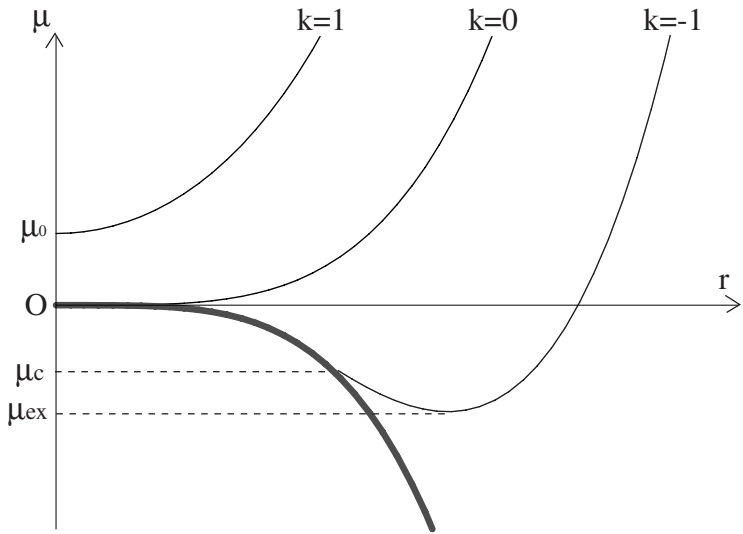}
\caption{\label{M-r}
The curves ${\mu}={\mu}_{\rm h}(r)$ and ${\mu}={\mu}_{\rm b}(r)$ in the GR branch with $\Lambda<0$, $\alpha>0$, and $1+4\alpha\Lambda/3 > 0$.
Thin curves correspond to ${\mu}={\mu}_{\rm h}(r)$ for each $k$.
A thick curve corresponds to ${\mu}={\mu}_{\rm b}(r)$.
The metric becomes complex and unphysical in the region of ${\mu}<{\mu}_{\rm b}(r)$.
The curve ${\mu}={\mu}_{\rm h}(r)$ with $k=-1$ terminates on the curve ${\mu}={\mu}_{\rm b}(r)$ at $\mu=\mu_c$.}
\end{center}
\end{figure}

The relation between the mass parameter $\mu$ and the radius of the Killing horizon $r_{\rm h}$ [defined by $h(r_{\rm h})=0$] is given by 
\begin{align}
\mu=16\alpha k^2+8kr_{\rm h}^2-\frac43 \Lambda r_{\rm h}^4=:\mu_{\rm h}(r_{\rm h}).
\end{align}
The relation between $\mu$ and $r_{\rm b}$ is given by
\begin{align}
\mu=-\frac{1}{\alpha }\biggl(1+\frac43\alpha\Lambda\biggl)r_{\rm b}^{4}=:\mu_{\rm b}(r_{\rm b}).
\end{align}
The number of horizons, their properties, and the existence of singularities depending on the parameters are understood from the functional forms of $\mu=\mu_{\rm h}(r)$ and $\mu=\mu_{\rm b}(r)$, which are drawn in Fig.~\ref{M-r}. 
(See~\cite{tm2005} for the detailed analysis.)
We summarize the global structures as follows.
For $k=1$, there is one nondegenerate outer horizon for $\mu>\mu_0:=16\alpha$ and no horizon for $\mu \le \mu_0$.
For $k=0$, there is one nondegenerate outer horizon for $\mu>0$ and no horizon for $\mu \le 0$.
For $k=-1$, there is one nondegenerate outer horizon for $\mu \ge  \mu_{\rm c}(:=-16\alpha (1+4\alpha\Lambda/3))$, two nondegenerate (outer and inner) horizons for $\mu_{\rm ex}(:=12 (1+4\alpha\Lambda/3)/\Lambda)<\mu<\mu_{\rm c}$, one degenerate horizon for $\mu=\mu_{\rm ex}$, and no horizon for $\mu<\mu_{\rm ex}$.

\begin{figure}[htbp]
\begin{center}
\includegraphics[width=0.5\linewidth]{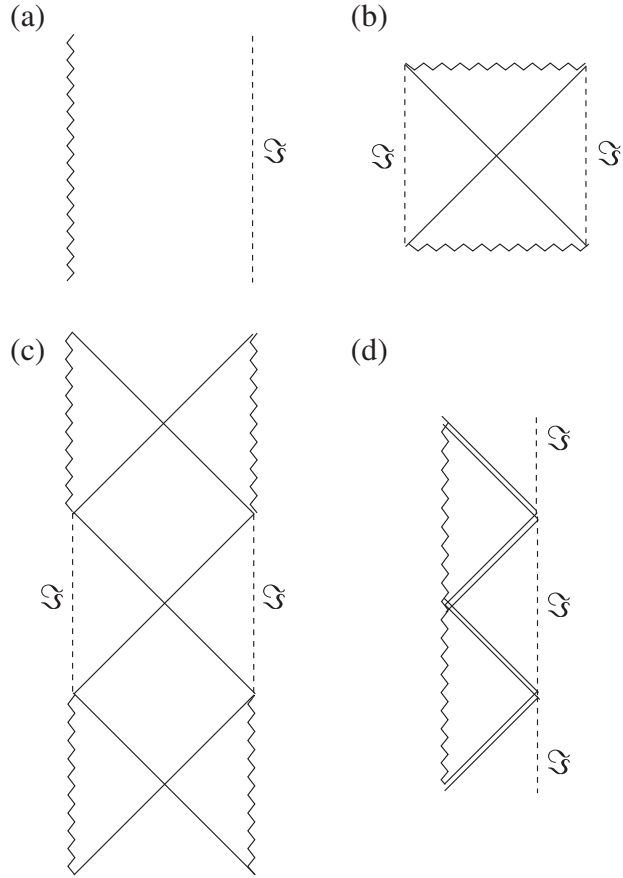}
\caption{\label{Penrose1}
The Penrose diagrams of the bulk spacetime (\ref{h-eq}) in the GR branch with $\alpha>0$, $\Lambda<0$, $1+4\alpha\Lambda/3 > 0$, and $\mu<0$.
A zig-zag line and a dashed line represent a branch singularity and infinity ($\Im$), respectively.
A thin line and a double line represent a nondegenerate and a degenerate Killing horizon, respectively.
The diagram (a) corresponds to the cases of $k=1$ and $0$.
For $k=-1$, the diagrams (b), (c), (d), and (a) correspond to the cases of $\mu_{\rm c} \le \mu<0$, $\mu_{\rm ex} < \mu<\mu_{\rm c}$,  $\mu=\mu_{\rm ex}$, and $\mu<\mu_{\rm ex}$, respectively.
}
\end{center}
\end{figure}

Near $r=0$, the metric function in the GR branch behaves as
\begin{align}
h(r)\simeq \biggl(k-\frac14\sqrt{\frac{\mu}{\alpha}}\biggl)+\frac{1}{4\alpha}r^2-\frac{1}{8\alpha^{3/2}\mu^{1/2}}\biggl(1+\frac43 \alpha\Lambda\biggl)r^4+O(r^8), \label{h-center}
\end{align}  
which is valid for $\mu>0$.
Therefore, the central singularity is timelike, null, and spacelike for $k-(1/4)\sqrt{\mu/\alpha}>0$, $k-(1/4)\sqrt{\mu/\alpha}=0$, and $k-(1/4)\sqrt{\mu/\alpha}<0$, respectively.
On the other hand, near the branch singularity, the metric function behaves as
\begin{align}
h(r)\simeq \biggl(k+\frac{r_{\rm b}^2}{4\alpha}\biggl)-\frac{r_{\rm b}^{3/2}}{2\alpha}\sqrt{1+\frac43 \alpha\Lambda}(r-r_{\rm b})^{1/2}. \label{h-branch}
\end{align}  
Therefore, the branch singularity is timelike and spacelike for $k+r_{\rm b}^2/(4\alpha)>0$ and $k+r_{\rm b}^2/(4\alpha) \le0$, respectively.
The Penrose diagrams in the case of $\mu<0$, which is of our interest, are drawn in Fig.~\ref{Penrose1}.

\subsection{Friedmann equation on the brane}
We consider a three-brane in the bulk spacetime (\ref{h-eq}), which is a timlike hypersurface described by $r=a(\tau)$ and $t=T(\tau)$, where the parameter $\tau$ is the proper time on the brane. 
The tangent vector to the brane is written as
\begin{equation}
u^\mu\frac{\partial}{\partial x^\mu}={\dot T} \frac{\partial}{\partial t}+{\dot a}\frac{\partial}{\partial r},
\end{equation}
where a dot denotes the differentiation with respect to $\tau$.
The normalization condition $u_\mu u^\mu=-1$ leads to
\begin{equation}
1=h(a){\dot T}^2-\frac{{\dot a}^2}{h(a)} \label{norm}
\end{equation}
and the induced metric of the three-brane ${\bar g_{ab}}$ is given by
\begin{equation}
ds_4^2={\bar g_{ab}}dy^ady^b= -d\tau^2+a(\tau)^2 \left[d\chi^2+f_k(\chi)^2(d\theta^2+\sin^2\theta d\phi^2) \right].\label{1stout}
\end{equation}
This is the FRW metric with the spatial curvature $k$.

The dynamics of the three-brane, namely the behavior of the scale factor $a(\tau)$ on the brane, is determined by the junction condition.
Here we simply assume the $Z_2$-symmetry of reflection with respect to the brane; we take two copies of the bulk spacetime with $r<a(\tau)$ and paste them at $r=a(\tau)$.
The junction condition in Einstein-Gauss-Bonnet gravity is given by~\cite{davis2003,GB-junction}
\begin{equation} 
[K^a_{~b}]_\pm-\delta^a_{~b}[K]_\pm+2\alpha\Bigr(3\varepsilon[J^a_{~b}]_\pm -\varepsilon\delta^a_{~b}[J]_\pm -2 P^a_{~dbf}[K^{df}]_\pm\Bigr)=-\varepsilon\kappa_5^2 S^a_{~b}, \label{j-condition} 
\end{equation}
where
\begin{align}
J_{ab} :=&{1\over 3} \left(2KK_{ad}K^{d}{}_{b}+K_{df}K^{df}K_{ab}-2K_{ad}K^{df}K_{fb}-K^2 K_{ab}\right), \\
P_{adbf}:=&R_{adbf}+2h_{a[f}R_{b]d}+2h_{d[b}R_{f]a} +Rh_{a[b}h_{f]d}.
\end{align}
$P_{adbf}$ is the divergence-free part of the Riemann tensor, i.e.,
\begin{equation}
D_a P^{a}{}_{dbf} \equiv 0,
\end{equation}
where $D_a$ is the covariant derivative on the brane.
We have introduced the notation
\begin{equation}
[X]_\pm:= X^+-X^-,
\end{equation}
where $X^\pm$ is the quantity $X$ evaluated either on the $+$ or $-$ side of the brane.
$\varepsilon=1$ and $\varepsilon=-1$ are used for timelike and spacelike branes, respectively.
For the dynamics of our three-brane, we take $\varepsilon=1$.
We consistently assume the form of the energy-momentum tensor $S^a{}_b$ on the brane as
\begin{equation}
S^a{}_b = \mbox{diag} (-\rho,p,p,p) + \mbox{diag}(-\sigma,-\sigma,-\sigma,-\sigma,),
\end{equation}
where $\rho$ and $p$ are the energy density and pressure of a perfect fluid on the brane, and the constant $\sigma$ is the brane tension. 

The junction condition (\ref{j-condition}) with $\varepsilon=1$ gives the modified Friedmann equation for the brane universe as
\begin{equation}
\frac{\kappa^4_5}{36}(\rho+\sigma)^2 = \left(\frac{h(a)}{a^2}+H^2\right)\left[1+\frac{4\alpha}{3} \left(\frac{3k- h(a)}{a^2}+2H^2\right)\right]^2, \label{GB-F}
\end{equation}
where $H := {\dot a}/a$~\cite{davis2003}.
The energy-conservation equation on the brane is obtained in the standard form as
\begin{equation}
{\dot \rho}=-3H(p+\rho). \label{em-cons}
\end{equation}
In order to close the system, we have to introduce an equation of state for matter.
We assume the following linear equation of state:
\begin{equation}
p=(\gamma-1)\rho. \label{eos}
\end{equation}
Equation~(\ref{em-cons}) is then integrated to give
\begin{equation}
\rho=\frac{\rho_0}{a^{3\gamma}}, \label{energy}
\end{equation}
where a constant $\rho_0$ is assumed to be positive, so that $\rho$ is a monotonically
decreasing function of $a$ for $\gamma>0$.
Because $\gamma=0$ is equivalent to the cosmological constant, we do not consider this case.
We assume $0<\gamma\le 2$ which satisfies the dominant energy condition.

In the Randall-Sundrum model, the Friedmann equation on the brane~\cite{krausida} becomes
\begin{equation}
H^2=\frac{\kappa^4_5}{36}\biggl(\frac{\rho_0}{a^{3\gamma}}+\sigma\biggl)^2-\frac{k}{a^2}+\frac{\mu}{8a^4}+\frac16 \Lambda =:V_{\rm GR}(a). \label{GR-F}
\end{equation}
Now we have a one-dimensional potential problem with one dynamical degree of freedom $a(\tau)$ and the qualitative behavior of $a$ is completely understood by the form of the potential $V_{\rm GR}$.
The region with $V_{\rm GR}(a)>0$ is the allowed region for dynamics.
The above equation (\ref{GR-F}) is compared with the standard four-dimensional Friedmann equation:
\begin{equation}
H^2=\frac{\kappa^2_4}{3}\frac{\rho_0}{a^{3\gamma}}-\frac{k}{a^2}+\frac13 \Lambda_4=:V_{\rm stand}(a), \label{F-eq}
\end{equation}
where $\kappa_4$ is defined by the four-dimensional gravitational constant $G_4$ as $\kappa_4:=\sqrt{8\pi G_4}$ and $\Lambda_4$ is the four-dimensional cosmological constant.

In the presence of the Gauss-Bonnet term, the dynamics of the brane is drastically modified.
Equation (\ref{GB-F}) in the GR branch is written as
\begin{align}
\frac{\kappa^4_5}{256\alpha^2}\biggl(\frac{\rho_0}{a^{3\gamma}}+\sigma\biggl)^2 =& \left(\frac{k}{a^2}+\frac{1-\sqrt{A}}{4\alpha}+H^2\right)\left(\frac{k}{a^2}+\frac{2+\sqrt{A}}{8\alpha}+ H^2\right)^2,\label{F-eqGR}\\
A:=&1+\frac{\alpha\mu}{a^4}+\frac43 \alpha\Lambda.\label{A}
\end{align}
We note that $A<1$ is satisfied if $\mu \le 0$.
The necessary conditions for the above equation to be physical are
\begin{align}
A\ge 0,  \qquad \frac{k}{a^2}+\frac{1-\sqrt{A}}{4\alpha}+H^2 \ge 0. \label{n1}
\end{align}
For our purpose, we rewrite Eq.~(\ref{F-eqGR}) in the form of $H^2=V_{\rm GB}(a)$.
Since Eq.~(\ref{F-eqGR}) is a cubic algebraic equation for $H^2$, there can be three real roots at most.
However, it is shown that there is only one positive real root at most.

The proof is given as follows.
For this purpose, we define a cubic function $F(x)$ as
\begin{equation}
F(x):=\left(\frac{k}{a^2}+\frac{1-\sqrt{A}}{4\alpha}+x\right)\left(\frac{8k\alpha}{3a^2}+\frac{2+\sqrt{A}}{3}+ \frac83 \alpha x\right)^2-\frac{\kappa^4_5}{36}\biggl(\frac{\rho_0}{a^{3\gamma}}+\sigma\biggl)^2.
\end{equation}
The Friedmann equation (\ref{F-eqGR}) is now $F(x)=0$ with $x=H^2$.
The dynamics is allowed only if $F(x)=0$ admits real and positive roots.
We show that $F(x)=0$ has only one real positive root at most.

We obtain
\begin{equation}
\frac{dF}{dx}=\frac{64\alpha^2}{3}\left(\frac{k}{a^2}+\frac{2+\sqrt{A}}{8\alpha}+x\right)\biggl(\frac{k}{a^2}+\frac{2-\sqrt{A}}{8\alpha}+x\biggl)
\end{equation}
and the solutions of the algebraic equation $dF/dx=0$ are $x=x_{\pm}~(x_+>x_-)$, where 
\begin{equation}
x_{\pm}:=\frac{-2 \pm \sqrt{A}}{8\alpha}-\frac{k}{a^2}.
\end{equation}
Since we obtain
\begin{align}
F(x_+)=&-\frac{A^{3/2}}{18\alpha}-\frac{\kappa^4_5}{36}\biggl(\frac{\rho_0}{a^{3\gamma}}+\sigma\biggl)^2,\\
F(x_-)=&-\frac{\kappa^4_5}{36}\biggl(\frac{\rho_0}{a^{3\gamma}}+\sigma\biggl)^2,
\end{align}
which satisfy $F(x_+)<F(x_-)<0$, it is concluded that there is one real positive root for $F(x)=0$ at most.
A sufficient condition for the existence of a real positive root is  $F(0)<0$.

For $\alpha>0$, Eq.~(\ref{F-eqGR}) is solved to give the following modified Friedmann equation on the brane:
\begin{align}
H^2=&V_{\rm GB(+)}(a), \label{GB-F2} \\
V_{\rm GB(+)}(a):=&\frac{1}{8\alpha}\biggl[-\frac{8k\alpha}{a^2}-2+\biggl\{A^{3/2}+256\alpha^3P^2+16\sqrt{2\alpha^3P^2\biggl(128\alpha^3P^2+A^{3/2}\biggl)}\biggl\}^{1/3} \nonumber \\
&+A\biggl\{A^{3/2}+256\alpha^3P^2+16\sqrt{2\alpha^3P^2\biggl(128\alpha^3P^2+A^{3/2}\biggl)}\biggl\}^{-1/3}\biggl], \label{V_GB}\\
P^2:=&\frac{\kappa^4_5}{256\alpha^2}\biggl(\frac{\rho_0}{a^{3\gamma}}+\sigma\biggl)^2.
\end{align}
Here $V_{\rm GB(+)}(a)$ denotes the effective potential for $\alpha>0$.
It is clear that inside both cubic and square roots in the potential are non-negative. 
The dynamics is allowed in the domain of $a$ where $V_{\rm GB(+)}(a)$ is real and positive and the conditions (\ref{n1}) are satisfied.
Indeed, the second necessary condition in Eq.~(\ref{n1}) is always satisfied.
Using the form of the potential $V_{\rm GB(+)}(a)$, we can write it in the following form:
\begin{align}
0 \le &\frac{k}{a^2}+\frac{1-\sqrt{A}}{4\alpha}+V_{\rm GB}(a) \nonumber \\
=&\frac{1}{8\alpha}\biggl\{A^{3/2}+256\alpha^3P^2+16\sqrt{2\alpha^3P^2\biggl(128\alpha^3P^2+A^{3/2}\biggl)}\biggl\}^{-1/3} \nonumber \\
&\times \biggl[-\sqrt{A}+\biggl\{A^{3/2}+256\alpha^3P^2+16\sqrt{2\alpha^3P^2\biggl(128\alpha^3P^2+A^{3/2}\biggl)}\biggl\}^{1/3}\biggl]^2=:W_+(a).\label{W+}
\end{align}
It is obvious that this condition holds only for $\alpha>0$.

The effective potential $V_{\rm GB(-)}(a)$ for $\alpha<0$ is different from $V_{\rm GB(+)}(a)$ and given by  
\begin{align}
V_{\rm GB(-)}(a):=&-\frac{1}{8\alpha}\biggl[\frac{8k\alpha}{a^2}+2+\biggl\{-A^{3/2}-256\alpha^3P^2+16\sqrt{2\alpha^3P^2\biggl(128\alpha^3P^2+A^{3/2}\biggl)}\biggl\}^{1/3} \nonumber \\
&+A\biggl\{-A^{3/2}-256\alpha^3P^2+16\sqrt{2\alpha^3P^2\biggl(128\alpha^3P^2+A^{3/2}\biggl)}\biggl\}^{-1/3}\biggl]. \label{V_GB-2}
\end{align}
With this potential, the second necessary condition in Eq.~(\ref{n1}) becomes
\begin{align}
0\le &\frac{k}{a^2}+\frac{1-\sqrt{A}}{4\alpha}+V_{\rm GB}(a) \nonumber \\
=&-\frac{1}{8\alpha}\biggl\{-A^{3/2}-256\alpha^3P^2+16\sqrt{2\alpha^3P^2\biggl(128\alpha^3P^2+A^{3/2}\biggl)}\biggl\}^{-1/3} \nonumber \\
&\times \biggl[\sqrt{A}+\biggl\{-A^{3/2}-256\alpha^3P^2+16\sqrt{2\alpha^3P^2\biggl(128\alpha^3P^2+A^{3/2}\biggl)}\biggl\}^{1/3}\biggl]^2.\label{W-}
\end{align}
Therefore, the second necessary condition holds for $\alpha<0$.

The condition that inside the square root in the potential is nonnegative is given by 
\begin{align}
128\alpha^3P^2+A^{3/2}=\frac{\alpha\kappa^4_5}{2}\biggl(\frac{\rho_0}{a^{3\gamma}}+\sigma\biggl)^2+\biggl(1+\frac{\alpha\mu}{a^4}+\frac43 \alpha\Lambda\biggl)^{3/2} \le 0. \label{condition1}
\end{align}
This may be violated for some $a$.
On the other hand, the condition that inside the cubic root in the potential is non-negative is given by
\begin{align}
A^{3/2}+256\alpha^3P^2\le  &16\sqrt{2\alpha^3P^2\biggl(128\alpha^3P^2+A^{3/2}\biggl)}.\label{condition2}
\end{align}
This is always satisfied under the condition (\ref{condition1}) because the left-hand side is nonpositive.
Since the equality in Eq.~(\ref{condition2}) cannot be satisfied, inside the cubic root in the potential is positive definite.

Let us see what happens if inside the square root in the potential becomes zero.
There are two possibilities for that, $P=0$ and $128\alpha^3P^2+A^{3/2}=0$.
It is shown that the derivative of the potential blows up only in the latter case.
Therefore, the equality in Eq.~(\ref{condition1}) corresponds to a curvature singularity where $a$ and ${\dot a}$ are finite but ${\ddot a}$ blows up.

\section{Bouncing Gauss-Bonnet braneworld}
The bouncing solution is characterized by the transition from the contracting phase (${\dot a}<0$) to the expanding one (${\dot a}>0$) of the universe.
As seen in the previous section, once we write the Friedmann equation in the form of $H^2=V(a)$, $V(a)>0$ is the allowed region for dynamics.
We assume that there is at least one domain of positive $a$ with $V(a)>0$ since there is no dynamical solution otherwise.
Then, the bounce (recollapse) in a conventional sense occurs at the lower (upper) bound of this domain $a=a_{\rm B}$ satisfying $V(a_{\rm B})=0$.
The evolution of the contracting universe in the domain $a>a_{\rm B}$ momentarily stops at $a=a_{\rm B}$ and then starts to expand in the domain $a>a_{\rm B}$.
Therefore, the sufficient condition for this conventional bounce is $V(0)<0$.
(Similarly, the sufficient condition for recollapse is $\lim_{a\to \infty}V(a)<0$.)

In this section, we discuss the possibility of the bounce by the asymptotic analysis for $a\to 0$.
The asymptotic behavior for $a\to \infty$ is presented in Appendix~B.
We first consider the possibility of the conventional bounce in the Randall-Sundrum braneworld and the Gauss-Bonnet braneworld with $\mu>0$.
We will see that, in comparison with them, the situation in the Gauss-Bonnet braneworld with $\mu<0$ is very different.
The primary reason is that $a<r_{\rm b}$ is not in the physical domain of $a$.
We will see what happens when the brane hits the branch singularity in the bulk.

\subsection{Conventional bounce condition}
In this subsection, we consider the condition for the bounce in a conventional sense.
In the standard Friedmann cosmology (\ref{F-eq}), we obtain
\begin{equation}
\lim_{a\to 0}V_{\rm stand}(a)\simeq \frac{\kappa^2_4}{3}\frac{\rho_0}{a^{3\gamma}}-\frac{k}{a^2}.
\end{equation}
Hence, the bounce occurs only for $k=1$ where $0<\gamma<2/3$ or $\gamma=2/3$ with $\rho_0<3/\kappa_4^2$ is satisfied.
In the case of  $k=1$, $\gamma=2/3$ and $\rho_0=3/\kappa_4^2$, the spacetime is just (A)dS for $\Lambda_4(<)>0$.

In the Randall-Sundrum braneworld (\ref{GR-F}), where we assume $\Lambda<0$ and $\mu \ne 0$, we obtain
\begin{align}
\lim_{a\to 0}V_{\rm GR}(a)\simeq& \frac{\kappa^4_5}{36}\frac{\rho_0^2}{a^{6\gamma}}+\frac{\mu}{8a^4}.
\end{align}
Hence, the bounce occurs for $0<\gamma<2/3$ with $\mu<0$.
For $\gamma=2/3$, the bounce occurs for $\mu<\mu_{\rm GR(\gamma=2/3)}$, where
\begin{align}
\mu_{\rm GR(\gamma=2/3)}:=-\frac{2\kappa^4_5\rho_0^2}{9}.
\end{align}
If a precise relation between $\rho_0$ and $\mu$ is satisfied as $\mu=\mu_{\rm GR(\gamma=2/3)}$, the bounce condition is more complicated for $\gamma=2/3$.
The bounce conditions are summarized in Table~\ref{table:bGR}.
\begin{table}[h]
\begin{center}
\caption{\label{table:bGR} Conditions for the conventional bounce in the Randall-Sundrum braneworld for arbitrary $k$. 
}
\begin{tabular}{c|@{\qquad}c@{\qquad}c@{\qquad}c}
\hline \hline
  & $0<\gamma<2/3$ & $\gamma=2/3$ & $2/3<\gamma \le 2$   \\\hline
$\mu>0$ & No bounce & No bounce & No bounce \\\hline
$\mu<0$ & Bounce & (No) Bounce for $\mu<(>)\mu_{\rm GR(\gamma=2/3)}$ & No bounce \\
\hline \hline
\end{tabular}
\end{center}
\end{table} 

In the Gauss-Bonnet braneworld (\ref{GB-F2}), the bounce condition is modified a lot.
Let us consider the case with $\mu>0$.
Our purpose is to clarify the asymptotic behavior of the potential for $a\to 0$.
For this purpose, the asymptotic form of $W_+(a)$ defined by Eq.~(\ref{W+}) is useful.
We write $W_+(a)$ in the following form:
\begin{align}
\frac{8\alpha a^2}{(\alpha\mu)^{1/2}}W_+(a)=&\sqrt{1+\biggl(1+\frac43\alpha\Lambda\biggl)\frac{a^4}{\alpha\mu}}\biggl\{1+256w^2+\frac{256}{\sqrt{128}}w\sqrt{1+128w^2}\biggl\}^{-1/3} \nonumber \\
&\times \biggl[1-\biggl\{1+256w^2+\frac{256}{\sqrt{128}}w\sqrt{1+128w^2}\biggl\}^{1/3}\biggl]^2=:L(w),\\
w:=&(\alpha^3A^{-3/2}P^2)^{1/2}.
\end{align}
$w$ is written in terms of $a$ as
\begin{align}
w=&\biggl[\frac{\rho_0^2\alpha\kappa^4_5a^{6-6\gamma}}{256(\alpha\mu)^{3/2}}\biggl\{1+\biggl(1+\frac43 \alpha\Lambda\biggl)\frac{a^4}{\alpha\mu}\biggl\}^{-3/2}\biggl(1+\frac{\sigma a^{3\gamma}}{\rho_0}\biggl)^2\biggl]^{1/2},
\end{align}
which is expanded around $a=0$ as
\begin{align}
\lim_{a\to 0}w \simeq &\frac{(\rho_0^2\alpha\kappa^4_5)^{1/2}a^{3-3\gamma}}{16(\alpha\mu)^{3/4}}\biggl\{1-\frac34\biggl(1+\frac43 \alpha\Lambda\biggl)\frac{a^4}{\alpha\mu}\biggl\}\biggl|1+\frac{\sigma a^{3\gamma}}{\rho_0}\biggl|.
\end{align}
The derivative of $w$ converges (diverges) for $0<\gamma \le 2/3$ ($2/3< \gamma \le 2$) near $a=0$ as
\begin{align}
\lim_{a\to 0}\frac{dw}{da} \simeq &\frac{(3-3\gamma)(\rho_0^2\alpha\kappa^4_5)^{1/2}a^{2-3\gamma}}{16(\alpha\mu)^{3/4}}.
\end{align}
Hence, $L(w)$ can be expanded around $a=0$ for $0<\gamma \le 2/3$ as
\begin{align}
\lim_{a\to 0}L(w)\simeq &\frac{512}{9}w^2 \simeq \frac{2\rho_0^2\alpha\kappa^4_5}{9(\alpha\mu)^{3/2}}a^{6-6\gamma},
\end{align}
from which we obtain 
\begin{align}
\lim_{a\to 0}W_+(a)\simeq & \frac{\rho_0^2\kappa^4_5}{36\alpha\mu}a^{4-6\gamma}.
\end{align}
Thus, asymptotic behavior of the potential around $a=0$ for $0<\gamma \le 2/3$ is 
\begin{align}
\lim_{a\to 0}V_{\rm GB(+)}(a)\simeq & -\frac{1}{a^2}\biggl(k-\frac{\sqrt{\alpha\mu}}{4\alpha}\biggl)-\frac{1}{4\alpha}+ \frac{\rho_0^2\kappa^4_5}{36\alpha\mu}a^{4-6\gamma}.
\end{align}
For  $2/3<\gamma <1$, we only know $\lim_{a\to 0}w=O(a^{\bar\delta})$ and $\lim_{a\to 0}L(a)=O(a^{\delta})$, where ${\bar\delta}$ and $\delta$ are positive constants.
Using them, we obtain
\begin{align}
\lim_{a\to 0}W_+(a)=&O(a^{\delta-2}),\\
\lim_{a\to 0}V_{\rm GB(+)}(a)=&-\frac{1}{a^2}\biggl(k-\frac{\sqrt{\alpha\mu}}{4\alpha}\biggl)-\frac{1}{4\alpha}+O(a^{\delta-2}),
\end{align}
from which we realize the leading term around $a=0$.
It is concluded from the analysis that, for $0<\gamma<1$, no bounce occurs for $k=0,-1$ or $k=1$ with $\mu>16\alpha (\equiv \mu_0)$, while the bounce occurs for  $k=1$ with $\mu<\mu_0$.
In the case of $k=1$ with $\mu= \mu_0$, the bounce occurs for $0<\gamma <2/3$, while a more careful analysis is required for $2/3<\gamma <1$.
For $\gamma=2/3$, the (no) bounce condition is given as $\mu>(<)\rho_0^2\kappa^4_5/9$.

For $1<\gamma \le 2$, $w$ blows up near $a=0$ and
\begin{align}
\lim_{a\to 0}V_{\rm GB(+)}(a)\simeq&\frac{(2\rho_0^2\alpha\kappa^4_5)^{1/3}}{8\alpha a^{2\gamma}}.
\end{align}
Therefore, the bounce does not occur.

For $\gamma=1$, $w$ behaves near $a=0$ as 
\begin{align}
\lim_{a\to 0}w \simeq &\frac{(\rho_0^2\alpha\kappa^4_5)^{1/2}}{16(\alpha\mu)^{3/4}}\biggl|1+\frac{\sigma a^{3}}{\rho_0}\biggl|.
\end{align}
Using this, we obtain
\begin{align}
\lim_{a\to 0}V_{\rm GB(+)}(a)\simeq &-\frac{1}{a^2}\biggl(k-\frac{\sqrt{\alpha\mu}}{4\alpha}\biggl)+\frac{(\alpha\mu)^{1/2}}{8\alpha a^2}\biggl\{1+256w_0^2+\frac{256}{\sqrt{128}}w_0\sqrt{1+128w_0^2}\biggl\}^{-1/3} \nonumber \\
&\times \biggl[1-\biggl\{1+256w_0^2+\frac{256}{\sqrt{128}}w_0\sqrt{1+128w_0^2}\biggl\}^{1/3}\biggl]^2-\frac{1}{4\alpha}+O(a), \nonumber \\
w_0:=&\frac{(\rho_0^2\alpha\kappa^4_5)^{1/2}}{16(\alpha\mu)^{3/4}}.
\end{align}
Hence, the bounce does not occur for $k=0,-1$ because the coefficient of $a^{-2}$ is positive.
On the other hand, the bounce occurs (does not occur) for $k=1$ if the coefficient of $a^{-2}$ is nonpositive (positive).
The result is summarized in Table~\ref{table:GB}.
\begin{table}[h]
\begin{center}
\caption{\label{table:GB} Conditions for the conventional bounce in the Gauss-Bonnet braneworld with $\mu>0$.
}
\begin{tabular}{c|c@{\quad}c@{\quad}c@{\quad}c}
\hline \hline
  &  $k=1$ & $k=0$ & $k=-1$    \\\hline	
$0< \gamma<2/3$ &  (No) bounce for $\mu \le(>)\mu_0$ & No bounce & No bounce  \\ \hline
$2/3\le  \gamma<1$ &  (No) bounce for $\mu<(>)\mu_0$ & No bounce & No bounce  \\ \hline
$\gamma=1$ &  See the maintext & No bounce & No bounce  \\ \hline
 $1<\gamma \le 2$ &  No bounce & No bounce & No bounce  \\
\hline \hline
\end{tabular}
\end{center}
\end{table} 

\subsection{Novel bouncing Gauss-Bonnet braneworld for $\mu<0$}
We have seen the bounce conditions for $\mu>0$ in the Gauss-Bonnet braneworld.
In this subsection, we discuss the case with $\mu<0$, where the situation is drastically changed.
As explained, $a<r_{\rm b}$ is not in the physical domain of $a$ and so we focus on the behavior of the potential around $a=r_{\rm b}$.
If the potential is non-negative, the brane reaches $a=r_{\rm b}$.
If the potential is negative near $a=r_{\rm b}$, the bounce (in the conventional sense) occurs or a singularity appears at some $a>r_{\rm b}$.

First we present the condition that the brane hits the branch singularity in the bulk spacetime.
The function $A$ defined by Eq.~(\ref{A}) behaves near $a=r_{\rm b}$ as
\begin{align}
A(a)\simeq &-\frac{4\alpha\mu}{r_{\rm b}^5}(a-r_{\rm b}).
\end{align}
The behavior of $V_{\rm GB(+)}(a)$ around $a=r_{\rm b}$ is given by
\begin{align}
V_{\rm GB(+)}(a)\simeq  &V_{\rm GB(+)}(r_{\rm b})+O(a-r_{\rm b}), 
\end{align}
where
\begin{align}
V_{\rm GB(+)}(r_{\rm b})= &-\frac{k}{r_{\rm b}^2}-\frac{1}{4\alpha}+ P(r_{\rm b})^{2/3}.
\end{align}
Therefore, the brane reaches the branch singularity if $V_{\rm GB(+)}(r_{\rm b})\ge 0$.
This condition is satisfied if 
\begin{align}
h(r_{\rm b})=k+\frac{r_{\rm b}^2}{4\alpha} \le 0.
\end{align}
Therefore, if the branch singularity is spacelike, which is realized only for $k=-1$, the brane reaches there.
If the branch singularity is timelike, the condition $V_{\rm GB(+)}(r_{\rm b})\ge 0$ is written as
\begin{align}
\sigma \ge \frac{16\alpha}{\kappa^2_5}\biggl(\frac{k}{r_{\rm b}^2}+\frac{1}{4\alpha}\biggl)^{3/2}-\frac{\rho_0}{r_{\rm b}^{3\gamma}}.\label{branch-condition}
\end{align}
It is noted that this condition is not so sensitive about $\gamma$ and realized even with positive pressure ($\gamma>1$).
On the other hand, if 
\begin{align}
\sigma < \frac{16\alpha}{\kappa^2_5}\biggl(\frac{k}{r_{\rm b}^2}+\frac{1}{4\alpha}\biggl)^{3/2}-\frac{\rho_0}{r_{\rm b}^{3\gamma}}
\end{align}
is satisfied, $a=r_{\rm b}$ is not in the physical domains of $a$ and hence the bounce occurs or a singularity appears at some $a>r_{\rm b}$ instead.

We saw the condition under which the brane hits the branch singularity in the bulk.
Let us see what happens then.
Now the important fact is that the curvature invariants on the brane do {\it not} blow up even when the brane approaches the branch singularity in the bulk.
In fact, the behavior of the scale factor $a(\tau)$ near the branch singularity is obtained as
\begin{eqnarray}
a(\tau) &\simeq& r_{\rm b}+a_1(\tau-\tau_{\rm b})+O((\tau-\tau_{\rm b})^2), \label{asymp-a} \\
a_1^2&:=&r_{\rm b}^2V_{\rm GB(+)}(r_{\rm b}),
\end{eqnarray}
where $\tau_{\rm b}$ is the cosmological time on the brane to reach the branch singularity.
This is the Taylor series around $\tau=\tau_{\rm b}$ and hence the curvature invariants are all finite around there.  
However, because the allowed domain of $a$ is $a \ge r_{\rm b}$, we must take the minus and plus signs of $a_1$ for $\tau<\tau_{\rm b}$ and $\tau>\tau_{\rm b}$, respectively.
As a result, the evolution near $\tau=\tau_{\rm b}$ represents the transition from the collapsing phase ($\tau<\tau_{\rm b}$) to the expanding phase ($\tau>\tau_{\rm b}$).
Thus, the bouncing universe is realized on the brane.

From the bulk point of view, this process is that the brane reaches the branch singularity and passes across it.
It is emphasized that the spacetime on the brane is not free of singularities then, but there appears just a shell-type instantaneous singularity.
In the generic case, the derivative of $a(\tau)$ (velocity) is not continuous and the metric on the brane is $C^0$ at $\tau=\tau_{\rm b}$.
The junction condition on the brane then shows that there is a matter distribution on the spacelike hypersurface $\tau=\tau_{\rm b}$ on the brane.
This means that a shell-type singularity appears instantaneously on the brane at $\tau=\tau_{\rm b}$ but it is rather harmless since it stems from the thin-shell approximation of the brane as well as the branch singularity. 
(In the next subsection, we will show that the branch singularity may be considered as a massive thin-shell.)
With a fine-tuning giving $a_1\equiv 0$, in contrast, the metric on the brane becomes analytic around $\tau=\tau_{\rm b}$ and there is no shell-type singularity.

Our claim is that the brane reaches the branch singularity and passes through it in the generic case and this process is the collision of the three-brane with the shell of branch singularity from the bulk point of view.
In order to support this claim, we show in the next section that the branch singularity is indeed harmless for a finite body moving radially.
Then, a next natural arises; what is the bulk spacetime on the other side of the branch singularity?
In order to answer to this question, we have to consider the extension of the bulk spacetime beyond $r=r_{\rm b}$.

\subsection{$C^0$ extension of the bulk beyond the branch singularity}
In this subsection, we discuss the extension of the bulk spacetime beyond the branch singularity.
It is seen in (\ref{h-eq}) that the metric and its inverse are finite at the branch singularity.
This implies that the spacetime can be extended beyond $r=r_{\rm b}$ in the $C^0$ sense.
The $C^0$ extension is not unique in general; however, thanks to the Birkhoff's theorem in the system~\cite{GBbrane-FRWequation}, there are only two candidates for the extended spacetime, namely the GR and non-GR branches of the vacuum solution (\ref{h-eq}).
Among these two, the GR branch should be chosen because of its dynamical stability.
In Fig.~3(a), the global structure of a $C^0$-extended spacetime is drawn as an example.
[It is the extension of the spacetime of Fig.~2(a) and the diagrams become different for the spacetimes of Figs.~2(b), (c), and (d).]
\begin{figure}[htbp]
\begin{center}
\includegraphics[width=0.7\linewidth]{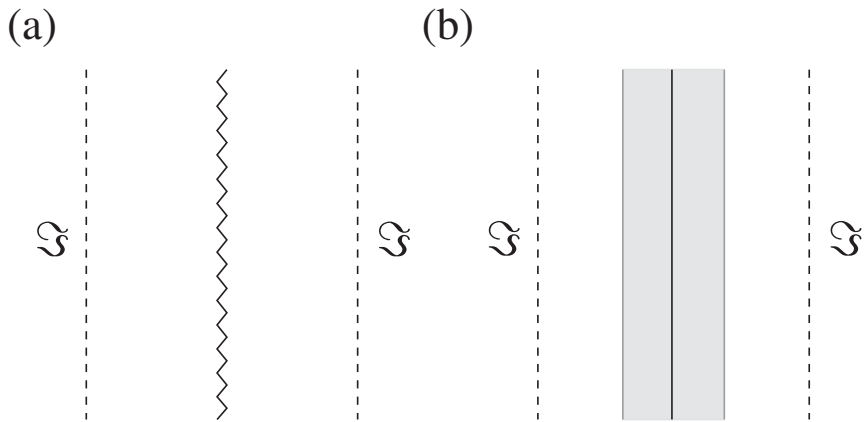}
\caption{\label{Penrose2}
The Penrose diagrams of (a) a $C^0$-extended bulk spacetime beyond the branch singularity [corresponding to Fig.~2(a)] and (b) its regularized version with a dust fluid for $k=-1$ and $\mu<\mu_{\rm ex}$.
In (b), a dust fluid fills in the shadowed region and the branch singularity is regularized and replaced by a regular wormhole throat (a thick solid line).
}
\end{center}
\end{figure}

We have constructed the $C^0$-extended bulk spacetime.
Now let us study the branch singularity in this spacetime in more detail.
The induced metric on $r=r_{\rm b}$ is given by
\begin{align}
ds_4^2=&{\bar g_{ab}}dy^ady^b \nonumber \\
=&  -h(r_{\rm b})dt^2+r_{\rm b}^2 \left[d\chi^2+f_k(\chi)^2(d\theta^2+\sin^2\theta d\phi^2) \right].\label{1stout-b}
\end{align}
We know that the derivatives of $h$ are not finite at $r=r_{\rm b}$.
The component of the extrinsic curvature $K_{ab}$ of the hypersurface $r=$constant diverges in the limit $r\to r_{\rm b}$.
Nevertheless, the junction condition provides a finite value of the energy-momentum tensor on $r=r_{\rm b}$.
This implies that the branch singularity can be identified as a massive shell in the extended bulk spacetime.
(It is noted that, in Einstein-Gauss-Bonnet gravity, even $C^0$ vacuum spacetimes can be constructed~\cite{gw2007}.)
We consider the cases where $r=r_{\rm b}$ is timelike ($h(r_{\rm b})>0$) and spacelike ($h(r_{\rm b}) < 0$), separately.
(A more careful treatment is required in the case of $h(r_{\rm b})=0$.)

First we consider the case where $r=r_{\rm b}$ is timelike.
We have 
\begin{equation}
h(r_{\rm b}) =k+\frac{r_{\rm b}^2}{4\alpha}(>0)
\end{equation}
and the tangent vector to a timelike hypersurface $r=$constant is written as
\begin{equation}
u^\mu\frac{\partial}{\partial x^\mu}=\frac{1}{\sqrt{h(r)}}\frac{\partial}{\partial t},
\end{equation}
which satisfies $u_\mu u^\mu=-1$.
The unit normal one-form to this hypersurface $n_\mu$ is given by
\begin{equation}
n_\mu dx^\mu =\frac{1}{\sqrt{h(r)}}dr,
\end{equation}
where $n_\mu u^\mu=0$ and $n_\mu n^\mu=1$ are satisfied.

The extrinsic curvature of this hypersurface is obtained from $K_{ab}:=
(\nabla_\nu n_{\mu})e^\mu_a e^\nu_b$, where $e^\mu_a := \partial x^\mu/\partial y^a$. We
have
\begin{equation}
e^0_ady^a =dt, \qquad e^1_ady^a = 0, \qquad e^i_ady^a = \delta^i_j dy^j,
\end{equation}
and the nonzero components of $K^a{}_b$ are 
\begin{equation}
K^t{}_t =\frac{1}{2\sqrt{h(r)}}h'(r), \qquad K^i{}_j = -\frac{1}{r}\sqrt{h(r)}\delta^i{}_j,
\end{equation}
where a prime denotes differentiation with respect to $r$.

Taking the limit $r\to r_{\rm b}$, $K^t{}_t$ blows up because $h'$ blows up there.
However, the junction condition (\ref{j-condition}) (with $\varepsilon=1$) shows that $S^a{}_b$ on the hypersurface remains finite for $r\to r_{\rm b}$ as
\begin{equation}
\lim_{r\to r_{\rm b}}S^a{}_b = \mbox{diag} (-\rho_{\rm b},p_{\rm b},p_{\rm b},p_{\rm b}),
\end{equation}
where $\rho_{\rm b}$ and $p_{\rm b}$ are interpreted as the energy density and pressure at $r=r_{\rm b}$.
Using the following fact,
\begin{eqnarray}
\lim_{r\to r_{\rm b}}\biggl\{h'(4\alpha h(r)-4\alpha k -r^2)\biggl\}\to - \frac{\mu}{2r_{\rm b}^3},\label{crucial}
\end{eqnarray}
we actually obtain
\begin{eqnarray}
\rho_{\rm b}=-16\alpha \frac{h(r_{\rm b})^{3/2}}{r_{\rm b}^3}, \quad p_{\rm b}=\frac{8h(r_{\rm b})r_{\rm b}^4-\mu}{2\sqrt{h(r_{\rm b})}r_{\rm b}^5}.
\end{eqnarray}
Thus, the branch singularity can be considered as a massive thin shell.
Since $\rho_{\rm b}<0$, the matter on $r=r_{\rm b}$ violates the weak energy condition.

Next we consider the case where $r=r_{\rm b}$ is spacelike, which is realized only for $k=-1$.
In this case, we consider the tangent vector to the (spacelike) hypersurface as
\begin{equation}
u^\mu\frac{\partial}{\partial x^\mu}=\frac{1}{\sqrt{-h(r)}}\frac{\partial}{\partial t},
\end{equation}
which satisfies $u_\mu u^\mu=1$.
The unit normal one-form to the (spacelike) hypersurface $n_\mu$ is then given by
\begin{equation}
n_\mu dx^\mu =-\frac{1}{\sqrt{-h(r)}}dr,
\end{equation}
where $n_\mu u^\mu=0$ and $n_\mu n^\mu=-1$ are satisfied.
The nonzero components of $K^a{}_b$ are 
\begin{equation}
K^t{}_t =-\frac{1}{2\sqrt{-h(r)}}h'(r), \qquad K^i{}_j = \frac{1}{r}\sqrt{-h(r)}\delta^i{}_j.
\end{equation}
Here we write the energy-momentum tensor $S^a{}_b$ on the (spacelike) hypersurface as
\begin{equation}
\lim_{r\to r_{\rm b}}S^a{}_b = \mbox{diag} (P_{\rm b(r)},P_{\rm b(t)},P_{\rm b(t)},P_{\rm b(t)}),
\end{equation}
where $P_{\rm b(r)}$ and $P_{\rm b(t)}$ are radial pressure and tangential pressure at $r=r_{\rm b}$, respectively.
Using Eq.~(\ref{crucial}),  the junction condition (\ref{j-condition}) with $\varepsilon=-1$ gives
\begin{eqnarray}
P_{\rm b(r)}=16\alpha \frac{(-h(r_{\rm b}))^{3/2}}{r_{\rm b}^3}, \quad P_{\rm b(t)}=\frac{8h(r_{\rm b})r_{\rm b}^4+\mu}{2\sqrt{-h(r_{\rm b})}r_{\rm b}^5}.
\end{eqnarray}
They are finite and hence the branch singularity may be considered as a massive spacelike thin shell.
Since this is a spacelike shell, it is difficult to discuss the energy condition for the matter field there.

\subsection{A simple regularized bulk model with matter}
We have seen that the branch singularity is weak and a three-brane may pass through it.
However, of course, the theory is no more valid around there in the bulk because it is a curvature singularity.
To make matters worse, it is a naked singularity in most cases and so it causes a problem of the boundary condition when we consider, for example,  the development of the perturbations in the bulk.
However, the weakness of the singularity could suggest that it is regularized in the effective theory of the final theory.
If so, the spacetime around the branch singularity should be described by some totally regular spacetime which approaches the vacuum spacetime in the far region.
Here the correction to the original theory around the singularity would appear in the field equation as an effective energy-momentum tensor, which might violate the energy conditions.

In this subsection, we present an exact bulk model with a simple matter field as an example of such spacetimes.
Using this solution as a bulk spacetime, we can obtain a more smooth bouncing universe on the brane. 
This simple construction works only for $k=-1$.

We consider a timelike dust fluid, of which energy-momentum tensor is given by 
\begin{eqnarray}
{T}_{\mu\nu}=\rho {\bar u}_{\mu}{\bar u}_{\nu},
\end{eqnarray}
where ${\bar u}^\mu$ and $\rho$ are the five-velocity of a fluid element and energy density, respectively.
For this matter field in Einstein-Gauss-Bonnet gravity, there is the following exact solution for $k \ne 0$~\cite{mn2008}:
\begin{align}
\label{met}
\D s^2&=-\D {\bar t}^2+\D {\bar x}^2+r({\bar x})^2\left[d\chi^2+f_k(\chi)^2(d\theta^2+\sin^2\theta d\phi^2) \right],\\
r&=\sqrt{\frac{3k}{\Lambda}}\cosh\biggl(\sqrt{-\frac{\Lambda}{3}}{\bar x}\biggl),\\
\rho&=\frac{\Lambda(3+4\alpha\Lambda)}{3\kappa_5^2},\\
{\bar u}^\mu\frac{\partial}{\partial x^\mu}&=\frac{\partial}{\partial {\bar t}}.
\end{align}  
The solution with $k=-1$ and $\Lambda<0$ represents a wormhole supported by a dust fluid with negative energy density, of which throat radius is given by 
\begin{align}
r=r_{\rm t}:=\sqrt{\frac{3}{|\Lambda|}}.
\end{align}
The quasilocal mass~\cite{maeda2006,GBMS} of this solution with $k=-1$ is obtained as 
\begin{align}
m=\frac{3(3+4\alpha\Lambda)V_{3}^{(-1)}}{4\Lambda\kappa_5^{2}}\cosh^{4}\biggl(\sqrt{-\frac{\Lambda}{3}}{\bar x}\biggl).
\end{align}
It is seen that the energy density of dust is negative, and hence, the weak energy condition is violated.
Also, the quasilocal mass is everywhere negative.

This solution can be attached to the exterior vacuum solution (\ref{h-eq}) in the GR branch with $k=-1$, $\Lambda<0$, and a certain negative mass parameter $\mu$ at $r=r_{\rm s}$, where $r_{\rm s}$ is the surface radius of the dust region~\cite{maeda2006}.
$r_{\rm s}>r_{\rm t}$ and $h(r_{\rm s})>0$ are required then.
In the resulting spacetime, there is no singularity at all.
The Penrose diagram of one regularized bulk spacetime is drawn in Fig.~3(b) as an example.
In summary, the branch singularity for $k=-1$ can be regularized by this matter field.
Unfortunately, this simple construction does not work in other cases.

\section{Weakness of the branch singularity}
\label{Sec:weak}
In this section, we show that the branch singularity is a weak singularity.
The physical consequence of this property is that a finite body may reach there safely.

There are several definitions of the strength of a singularity.
(See~\cite{Joshi,Clarketext}.)
We first present the definition by Tipler~\cite{Tipler}.
Let ${\bar \gamma}: [\lambda_0,\lambda_{\rm s})\to M$ be an affinely parametrized causal geodesic which approaches a singularity as $\lambda \to \lambda_{\rm s}^-$, where $\lambda$ is an affine parameter.
Define $J_{\lambda_1}({\bar \gamma})$ for $\lambda_1\in [\lambda_0,\lambda_{\rm s})$ to be a set of maps $Z_{(I)}:[\lambda_0,\lambda_{\rm s})\to TM$ ($TM$ means the tangent bundle and $I=1,2,3,4~(I=1,2,3)$ for timelike (null) ${\bar \gamma}$) satisfying the following four:
\begin{align}
Z_{(I)}^\mu(\lambda)\in& T_{{\bar \gamma}(\lambda)}M,\\
Z_{(I)}^\mu(\lambda_1)=&0,\\
{\ddot Z}_{(I)}^\mu=&-R^\mu_{~~\nu\rho\sigma}Z_{(I)}^{\rho}k^\nu k^\sigma,\label{jacobi}\\
{Z}_{(I)}^\mu k_\mu=&0.
\end{align}  
where $k^\mu$ is the tangent of ${\bar \gamma}$.
Equation~(\ref{jacobi}) is called the Jacobi equation (or geodesic deviation equation).
Along a timelike geodesic, four independent Jacobi fields define a volume element $V(\lambda)$ along ${\bar \gamma}$ by the exterior product.
Along a null geodesic, three such fields define an area element which we also denote $V(\lambda)$.
A singularity is called {\it Tipler strong} if 
\begin{align}
\lim_{\lambda\to \lambda_{\rm s}^-}\inf V(\lambda)=0
\end{align}  
is satisfied for all $\lambda_1\in [\lambda_0,\lambda_{\rm s})$ and all four (three) linearly independent Jacobi fields $Z \in J_{\lambda_1}({\bar \gamma})$~\cite{Tipler}.
The singularity is called Tipler weak if it is not Tipler strong.
This definition of the Tipler strong singularity intuitively says that any object that hits a strong singularity is crushed to zero volume (area).

The above definition ignores the case where $V(\lambda)$ blows up in the approach to the singularity.
Also, $V(\lambda)$ may remain finite overall when some of the elements of $J_{\lambda_1}({\bar \gamma})$ blow up but some others converge to zero.
These possibilities were pointed out by Nolan~\cite{strength} and Ori~\cite{ori2000}.
In order to include such situations, Ori defined deformationally strong singularity~\cite{ori2000}.
A singularity is called {\it deformationally strong} if it is either (i) Tipler strong, or (ii) if there exists an element of  $J_{\lambda_1}({\bar \gamma})$ that has infinite norm for $\lambda\to \lambda_{\rm s}^-$ for all $\lambda_1\in [\lambda_0,\lambda_{\rm s})$~\cite{ori2000}.
A singularity is called deformationally weak if it is not deformationally strong.

A singularity is Tipler weak if it is deformationally weak.
Here we show that the branch singularity is deformationally weak along radial causal geodesics.
The first task for this purpose is to clarify the asymptotic behavior of causal geodesics near the singularity.

\subsection{Asymptotic behavior of geodesics}
For the metric (\ref{h-eq}), the Lagrangian for a geodesic ${\bar \gamma}$ is written as
\begin{align}
L=&\frac12g_{\mu\nu}{\dot x}^\mu{\dot x}^\nu \\
=&-\frac12 h(r){\dot t}^2+\frac12 h(r)^{-1}{\dot r}^2+\frac12 r^2 \left[{\dot \chi}^2+f_k(\chi)^2({\dot \theta}^2+\sin^2\theta {\dot \phi}^2) \right],
\end{align}  
where a dot denotes the derivative with respect to the affine parameter $\lambda$.
Here $k^\mu:={\dot x^\mu}$ is a tangent vector of ${\bar \gamma}$ satisfying
\begin{align}
k^\mu k_\mu=\epsilon,\label{1stint}
\end{align}  
where $\epsilon$ is given as $0$ and $-1$ for a null and timelike geodesic, respectively.
The metric (\ref{h-eq}) does not depend on $t$ and $\phi$, so that from the Lagrange equation we find two independent conserved quantities along a geodesic:
\begin{align}
E&:= - \frac{\partial L}{\partial {\dot t}}=h(r){\dot t}, \label{ce1} \\
\Phi&:= \frac{\partial L}{\partial {\dot \phi}}=r^2f_k^2\sin^2\theta{\dot \phi}. \label{ce2}
\end{align}  
Hereafter we consider only radial geodesics and set the angler coordinates $\chi$, $\theta$, and $\phi$ all constants, which gives $\Phi=0$.
Then, the Lagrange equation for $r$ is given as
\begin{align}
{\ddot r}=\frac{\epsilon}{2} h', \label{geo2}
\end{align}  
of which the first integral is given from Eq.~(\ref{1stint}) as 
\begin{align}
{\dot r}^2=E^2+\epsilon h(r). \label{geo1}
\end{align}

Now we obtain the asymptotic behavior of the geodesic close to the singularity.
First we consider the case of timelike ${\bar \gamma}$ ($\epsilon=-1$).
Using the asymptotic expansion (\ref{h-center}), Eq.~(\ref{geo1}) gives the asymptotic behavior of $r(\lambda)$ near $r=0$ as
\begin{align}
r(\lambda)\simeq& {\bar r}_1(\lambda-\lambda_{\rm s})+{\bar r}_2(\lambda-\lambda_{\rm s})^{3},\label{ex-0}\\
{\bar r}_1^2:=&E^2+\epsilon \biggl(k-\frac14\sqrt{\frac{\mu}{\alpha}}\biggl),\\
{\bar r}_2:=&\frac{\epsilon }{24\alpha}{\bar r}_1.
\end{align}  
$\lambda=\lambda_{\rm s}$ is the affine time corresponding to $r=0$.

On the other hand, using the asymptotic expansion (\ref{h-branch}), the asymptotic behavior of $r(\lambda)$ near the branch singularity $r=r_{\rm b}$ is given as
\begin{align}
r(\lambda)\simeq&r_{\rm b}+ r_1(\lambda-\lambda_{\rm s}),\label{ex-b}\\
r_1^2:=&E^2+\epsilon \biggl(k+\frac{r_{\rm b}^2}{4\alpha}\biggl).
\end{align}  
Here $\lambda=\lambda_{\rm s}$ is the affine time corresponding to $r=r_{\rm b}$.

In the case of null geodesics ($\epsilon=0$), Eq.~(\ref{geo1}) gets simplified and is exactly solved to give
\begin{align}
{r}(\lambda)=r_{\rm b}\pm E(\lambda-\lambda_{\rm s}).
\end{align}  
We set $r_0=r_{\rm b}$ and $r_0=0$ for geodesics terminating in or emanating from the branch singularity and the central singularity, respectively.

\subsection{Strength of the branch singularity}
We analyze the strength of the branch singularity.
We adopt the similar method in four dimensions~\cite{strength}. 
(See~\cite{as2010} for the higher-dimensional analysis.)
The tangent vector to ${\bar \gamma}$ is given by 
\begin{align}
k^\mu\frac{\partial}{\partial x^\mu}&={\dot t}(\lambda)\frac{\partial}{\partial t}+{\dot r}(\lambda)\frac{\partial}{\partial r} \nonumber \\
&=\frac{E}{h(r)}\frac{\partial}{\partial t}+{\dot r}(\lambda)\frac{\partial}{\partial r},
\end{align}  
where $r(\lambda)$ satisfies Eq.~(\ref{geo1}).
A set of Jacobi fields $Z_{(I)}^\mu$ along ${\bar \gamma}$ satisfies the Jacobi equation (\ref{jacobi}).
Because the Jacobi equation is a linear equation for $Z_{(I)}^\mu$, a basis for the Jacobi fields can be found by obtaining all independent Jacobi fields in the radial two-space and in the tangential three-space.

In the tangent three-space, there are three independent parallel propagated (namely $k^\nu \nabla_\nu \eta_{(I)}^\mu=0$) unit spacelike vectors:
\begin{align}
\eta_{(2)}^\mu\frac{\partial}{\partial x^\mu}&=\frac{1}{r}\frac{\partial}{\partial \chi}, \label{space2} \\
\eta_{(3)}^\mu\frac{\partial}{\partial x^\mu}&=\frac{1}{rf_k(\chi)}\frac{\partial}{\partial \theta},\label{space4}\\
\eta_{(4)}^\mu\frac{\partial}{\partial x^\mu}&=\frac{1}{rf_k(\chi)\sin\theta}\frac{\partial}{\partial \phi}.\label{space4}
\end{align}  
Their dual one-forms $e^{(I)}_\mu dx^\mu$ (satisfying $e^{(I)}_\mu\eta_{(I)}^\mu=1$) are given by 
\begin{align}
e^{(2)}_\mu dx^\mu&=rd\chi,\label{space2-2} \\
e^{(3)}_\mu dx^\mu&=rf_k(\chi)d \theta,\label{space3-2}\\
e^{(4)}_\mu dx^\mu&=rf_k(\chi)\sin\theta d\phi.\label{space4-2}
\end{align}  
Hereafter, we discuss the cases where ${\bar \gamma}$ is timelike and null, separately.

\subsubsection{Along radial timelike geodesics}
We first consider the case where ${\bar \gamma}$ is timelike.
The dual one-form $e^{(0)}_\mu dx^\mu $ of $k^\mu(\partial/\partial x^\mu)$ (satisfying $e^{(0)}_\mu k^\mu=-1$) is given by 
\begin{align}
e^{(0)}_\mu dx^\mu&=-h{\dot t}dt+h^{-1}{\dot r}dr.
\end{align}  
A spacelike unit vector in the radial two-space orthogonal to $k^\mu(\partial/\partial x^\mu)$ is given by 
\begin{align}
\eta_{(1)}^\mu\frac{\partial}{\partial x^\mu}&=\frac{{\dot r}}{h(r)}\frac{\partial}{\partial t}+{\dot t}h(r)\frac{\partial}{\partial r} \nonumber \\
&=\frac{{\dot r}}{h}\frac{\partial}{\partial t}+E\frac{\partial}{\partial r},
\end{align}  
which is parallel propagated along ${\bar \gamma}$.
Its dual one-form $e^{(1)}_\mu dx^\mu$ is given by 
\begin{align}
e^{(1)}_\mu dx^\mu&=-{\dot r}dt+{\dot t}dr.
\end{align}  
We consider the following Jacobi fields along ${\bar \gamma}$:
\begin{align}
Z^\mu_{(I)}\frac{\partial}{\partial x^\mu}:=l_{(I)}(\lambda) \eta_{(I)}^\mu\frac{\partial}{\partial x^\mu} \quad (I=1,2,3,4),
\end{align}  
of which the norm is given by $l_{(I)}^2$.
The dual one-forms ${\hat e}^{(I)}_\mu$ of $Z_{(I)}~(I=1,2,3,4)$ are given by ${\hat e}^{(I)}_\mu:=l_{(I)} e^{(I)}_\mu$.
If ${\bar \gamma}$ is timelike, they define a volume four-form by exterior product and its norm is given by 
\begin{align}
V(\lambda)=|l_{(1)}||l_{(2)}||l_{(3)}|l_{(4)}|.
\end{align}  
The behavior of $l_{(I)}(\lambda)$ is determined by the Jacobi equation (\ref{jacobi}).

The Jacobi equation for $I=2,3,4$ gives
\begin{align}
0={\ddot l}_{(I)}- \frac{\epsilon h'}{2r}l_{(I)},\label{l-eq}
\end{align}  
where we used Eq.~(\ref{geo1}) and the fact that $\eta_{(I)}^\mu$ is parallel propagated.
Using Eq.~(\ref{geo2}), Eq.~(\ref{l-eq}) can be also written as
\begin{align}
0={\ddot l}_{(I)}-\frac{{\ddot r}}{r}l_{(I)}. \label{eq-vI}
\end{align}  
This equation is solved to give
\begin{align}
l_{(I)}(\lambda)=r(\lambda)\int^{\lambda}_{\lambda_1}\frac{d{\bar\lambda}}{r({\bar\lambda})^2},
\end{align}  
which satisfies $l_{(I)}(\lambda_1)=0$.
Since $r$ is finite for $\lambda=[\lambda_0,\lambda_{\rm s})$, $l_{(I)}(\lambda)$ is nonzero finite for $\lambda \to \lambda_{\rm s}^{-}$.

For $I=1$, the Jacobi equation gives
\begin{align}
0=&{\ddot Z}_{(1)}^t+R^t_{~~rtr}Z_{(1)}^{t}k^r k^r+R^t_{~~r r t}Z_{(1)}^{r}k^r k^t \nonumber \\
=&h^{-1}{\dot r}\biggl({\ddot  l_{(1)}}-\frac12 \epsilon  h''l_{(1)}\biggl), \\
0=&{\ddot Z}_{(1)}^r+R^r_{~~t t r}Z_{(1)}^{t}k^t k^r+R^r_{~~t r t}Z_{(1)}^{r}k^t k^t \nonumber \\
=&E\biggl({\ddot  l_{(1)}}-\frac12 \epsilon  h''l_{(1)}\biggl),
\end{align}  
where we used the fact that $\eta_{(1)}^\mu$ is parallel propagated.
Hence, both components give the same equation:
\begin{align}
0=&{\ddot  l_{(1)}}-\frac12 \epsilon  h''l_{(1)} \nonumber \\
=&{\ddot  l_{(1)}}-\frac{\dddot r}{{\dot r}}l_{(1)},
\end{align}  
where we used the derivative of Eq.~(\ref{geo2}) at the last equality.
Using Eq.~(\ref{ex-b}), we obtain the asymptotic solution around the branch singularity as
\begin{align}
l_{(1)}(\lambda)\simeq l_{(1)0}-\frac{\epsilon l_{(1)0} r_{\rm b}^{3/2}}{4\alpha r_1^{3/2}}\sqrt{1+\frac43 \alpha\Lambda} (\lambda-\lambda_{\rm s})^{1/2},
\end{align}  
where $l_{(1)0}$ is a nonzero constant and hence $l_{(1)}(\lambda)$ is nonzero finite for $\lambda\to \lambda_{\rm s}^-$.
Since all $l_{(I)}(\lambda)$ are nonzero finite for $\lambda\to \lambda_{\rm s}^-$, the branch singularity is deformationally weak along radial timelike geodesics.

\subsubsection{Along radial null geodesics}
Next we consider the case where ${\bar \gamma}$ is null.
In this case, the Jacobi fields only in the tangent space are relevant in order to define the strength of the singularity.
The strength is defined by the norm of the area three-form constructed by ${\hat {\bf e}}^{(I)}~(I=2,3,4)$:
\begin{align}
A(\tau)=|l_{(2)}||l_{(3)}|l_{(4)}|.
\end{align}

From Eq.~(\ref{l-eq}), the Jacobi equation for $I=2,3,4$ is integrated to give
\begin{align}
l_{(I)}(\lambda)=& l_{(I)1}(\lambda-\lambda_1),
\end{align}  
where $l_{(I)1}$ is a nonzero constant.
This satisfies $l_{(I)}(\lambda_1)=0$ and remains finite for $\lambda\to \lambda_{\rm s}^-$.
Hence, the branch singularity is deformationally weak also along radial null geodesics.

\section{Summary and discussions}
In this paper, we have presented a novel scenario for the bouncing universe in the five-dimensional braneworld in the framework of Einstein-Gauss-Bonnet gravity.
In this scenario, the branch singularity located at the finite physical radius in the bulk, which appears for the negative mass parameter,  plays an essential role.
We have shown that the branch singularity can be considered as a massive thin shell and a three-brane may pass through it.
As a result, the bouncing universe is realized on the brane.

The bulk spacetime is extended beyond the branch singularity in the $C^0$ sense and then the branch singularity is identified as a massive thin shell.
From the bulk point of view, the three-brane collides with another shell of branch singularity and continues to evolve into the extended spacetime.
From the brane point of view, the moment of the collision is the moment of the bounce.
This is not the conventional bounce because the derivative of the scale factor is discontinuous at the bounce moment.
Therefore, there appears a shell-type instantaneous singularity on the brane but the curvature invariants never blow up before or after then. 
This claim is strongly supported by the fact that the branch singularity is radially deformationally weak, which implies that the singularity is harmless for a finite body moving radially.
The present result opens a completely new possibility to achieve the bouncing brane universe as an effect of the higher-curvature terms.
Our scenario is not sensitive about the equation of state for the matter on the brane and does work even with ordinary matter.

Here we should make a brief comment on the anisotropy of the universe.
Although the assumption of the isotropy of the universe is justified at the recombination era by the distribution of the cosmic microwave background in the WMAP-Planck data, its validity is a nontrivial problem in the very early universe before that.
 In the standard FRW cosmology, it is expected by the BKL (Belinsky-Khalatnikov-Lifshitz) conjecture that such an assumption is no more valid and the behavior of the generic spacetime close to the initial singularity is quite anisotropic~\cite{we1997}.
In contrast, this claim is not necessarily true in the Randall-Sundrum braneworld.
It has been shown within the Bianchi class anisotropic cosmological models that the brane universe is generically isotropic close to the initial singularity. (See~\cite{BianchiBrane} for review.)
However, the similar analysis has not been done in the Gauss-Bonnet braneworld so far.
This is one of the important future subjects to clarify the validity of our bouncing scenario.

Our result also gives a suggestion on what would happen if the brane approaches the inner Cauchy horizon in the generalized Reissner-Nordstr\"om-AdS bulk spacetime with a U(1) gauge field in the Randall-Sundrum braneworld~\cite{bounce-U(1),hm2003}.
The rigorous analysis in the four-dimensional asymptotically flat case suggests that the mass inflation instability transforms the inner horizon into a weak singularity.
More precisely, the resulting singularity appears at a finite physical radius where the metric and its inverse are finite~\cite{massinflation}.
Therefore, the $C^0$ extension is possible beyond this singularity.
Then a natural question is whether the brane can reach and pass through it or not.
Of course, the appearance of the singularity means that the theory is no more reliable around there.
However, the result in the present paper suggests that the three-brane could safely pass through the inner horizon in the bulk.

Since the existence of the branch singularity stems from the quadratic nature of the theory, such a singularity is characteristic and must be quite generic in higher-curvature theories.
In Lovelock higher-curvature gravity~\cite{lovelock}, which contains general relativity and Einstein-Gauss-Bonnet gravity as special cases, such a singularity appears in this class of vacuum solutions rather generically~\cite{mwr2011}.
Interestingly, the central singularity is totally absent and the branch singularity is generic independent of the mass paraeter $\mu$ if there is a U(1) gauge field in the bulk spacetime in Einstein-Gauss-Bonnet gravity~\cite{GBBH-charge,tm2005b}.
Undoubtedly, the effect of such singularities in cosmology is an interesting problem and should be investigated further.

\subsection*{Acknowledgements}
The author thanks S.~Willison, Y.~Shtanov, and V.~Sahni for useful comments and discussions. 
He also thanks F.~Canfora for comments on the asymptotic analysis. 
This work has been funded by the Fondecyt Grants No. 1100328 and No. 1100755 and by the Conicyt grant "Southern Theoretical Physics Laboratory" ACT-91. 
This work was also partly supported by the JSPS Grant-in-Aid for Scientific Research (A) (22244030).
The Centro de Estudios Cient\'{\i}ficos (CECs) is funded by the Chilean Government through the Centers of Excellence Base Financing Program of Conicyt.

\appendix

\section{Geometric quantities}
For the metric (\ref{h-eq}), the nonzero components of the Christoffel symbol are given by 
\begin{align}
\Gamma^r_{~~tt}&=\frac12 hh',\quad \Gamma^t_{~~tr}=\frac{h'}{2h}, \quad \Gamma^r_{~~rr}=-\frac{h'}{2h},\\
\Gamma^\chi_{~~r\chi}&=\Gamma^\theta_{~~r\theta}=\Gamma^\phi_{~~r\phi}=\frac{1}{r},\\
\Gamma^r_{~~\chi\chi}&=-rh,\quad \Gamma^\theta_{~~\chi\theta}=\Gamma^\phi_{~~\chi\phi}=\frac{\partial_\chi f_k}{f_k},\\
\Gamma^r_{~~\theta\theta}&=-rhf_k^2,\quad \Gamma^\chi_{~~\theta\theta}=-f_k\partial_\chi f_k,\\
\Gamma^\phi_{~~\theta\phi}&=\frac{\cos\theta}{\sin\theta},\quad  \Gamma^r_{~~\phi\phi}=-rhf_k^2\sin^2\theta,\\
\Gamma^\chi_{~~\phi\phi}&=-f_k(\partial_\chi f_k)\sin^2\theta,\quad  \Gamma^\theta_{~~\phi\phi}=-\sin\theta\cos\theta,
\end{align}  
where a prime denotes the derivative with respect to $r$.
The nonzero components of the Riemann tensor are given by 
\begin{align}
R^t_{~~rtr}&=-\frac{h''}{2h},\quad R^r_{~~ttr}=-\frac12 hh'',\\
R^t_{~~\chi t\chi}&=-\frac12 rh',\quad R^t_{~~\theta t\theta}=-\frac12 rh'f_k^2,\quad  R^t_{~~\phi t\phi}=-\frac12 rh'f_k^2\sin^2\theta,\\
R^r_{~~\chi r \chi}&=-\frac12 rh' \quad R^r_{~~\theta r \theta}=-\frac12 rh'f_k^2  \quad R^r_{~~\phi r \phi}=-\frac12 rh'f_k^2\sin^2\theta,\\
R^\chi_{~~tt\chi}&=R^\theta_{~~tt\theta}=R^\phi_{~~tt\phi}=-\frac{hh'}{2r},\\
R^\chi_{~~rr\chi}&=R^\theta_{~~rr\theta}=R^\phi_{~~rr\phi}=\frac{h'}{2rh},\\
R^\chi_{~~\theta\chi\theta}&=f_k^2(k-h),\quad R^\chi_{~~\phi\chi\phi}=f_k^2(k-h)\sin^2\theta,\\
R^\theta_{~~\chi\chi\theta}&=R^\phi_{~~\chi\chi\phi}=-k+h,\\
R^\theta_{~~\phi\theta\phi}&=f_k^2(k-h)\sin^2\theta,\quad R^\phi_{~~\theta\theta\phi}=-f_k^2(k-h).
\end{align}

\section{Asymptotic behavior for $a\to \infty$}
In this Appendix, we discuss the asymptotic behavior of the brane universe for $a\to \infty$.
In the standard Friedmann cosmology (\ref{F-eq}), we obtain
\begin{equation}
\lim_{a\to \infty}V_{\rm stand}(a)= \frac13\Lambda_4.
\end{equation}
Hence, the recollapse occurs for $\Lambda_4 < 0$.

In the Randall-Sundrum braneworld (\ref{GR-F}), we obtain
\begin{align}
\lim_{a\to \infty}V_{\rm GR}(a)\simeq \frac{\kappa^4_5}{36}\sigma^2+\frac16 \Lambda+\biggl(\frac{\kappa^4_5}{18}\frac{\sigma\rho_0}{a^{3\gamma}}-\frac{k}{a^2}\biggl)+\frac{\kappa^4_5}{36}\frac{\rho_0^2}{a^{6\gamma}}+\frac{\mu}{8a^4}.
\end{align}
Hence the recollapse condition is given by
\begin{align}
\kappa^4_5\sigma^2<- 6\Lambda.\label{recollapse-gr}
\end{align}
If a precise relation between $\Lambda$ and $\sigma$ is satisfied as $\Lambda=-\kappa^4_5\sigma^2/6$, the condition becomes more complicated.

In the Gauss-Bonnet braneworld with $\alpha>0$, the recollapse condition is modified as
\begin{align}
\lim_{a\to \infty}V_{\rm GB(+)}(a) =&\frac{1}{8\alpha}\biggl[-2+\biggl\{\biggl(1+\frac43\alpha\Lambda\biggl)^{3/2}+\alpha\sigma^2\kappa^4_5+\alpha\sigma^2\kappa^4_5\sqrt{1+\frac{2}{\alpha\sigma^2\kappa^4_5}\biggl(1+\frac43\alpha\Lambda\biggl)^{3/2}}\biggl\}^{1/3} \nonumber \\
&+\biggl(1+\frac43\alpha\Lambda\biggl)\biggl\{\biggl(1+\frac43\alpha\Lambda\biggl)^{3/2}+\alpha\sigma^2\kappa^4_5+\alpha\sigma^2\kappa^4_5\sqrt{1+\frac{2}{\alpha\sigma^2\kappa^4_5}\biggl(1+\frac43\alpha\Lambda\biggl)^{3/2}}\biggl\}^{-1/3}\biggl] \nonumber \\
&<0.\label{recollapse-gb}
\end{align}
On the other hand, in the case of $\alpha<0$, we obtain
\begin{align}
\lim_{a\to \infty}V_{\rm GB(-)}(a)=&-\frac{1}{8\alpha}\biggl[2+\biggl\{-\biggl(1+\frac43\alpha\Lambda\biggl)^{3/2}-\alpha\sigma^2\kappa^4_5-\alpha\sigma^2\kappa^4_5\sqrt{1+\frac{2}{\alpha\sigma^2\kappa^4_5}\biggl(1+\frac43\alpha\Lambda\biggl)^{3/2}}\biggl\}^{1/3} \nonumber \\
&+\biggl(1+\frac43\alpha\Lambda\biggl)\biggl\{-\biggl(1+\frac43\alpha\Lambda\biggl)^{3/2}-\alpha\sigma^2\kappa^4_5-\alpha\sigma^2\kappa^4_5\sqrt{1+\frac{2}{\alpha\sigma^2\kappa^4_5}\biggl(1+\frac43\alpha\Lambda\biggl)^{3/2}}\biggl\}^{-1/3}\biggl] \nonumber \\
&<0.\label{recollapse-gb2}
\end{align}
The above limit is not real if inside the square root is negative.
This condition is given by 
\begin{align}
\sigma^2> -\frac{2}{\alpha\kappa^4_5}\biggl(1+\frac43 \alpha\Lambda\biggl)^{3/2}.
\end{align}
If the above inequality is satisfied, there is an upper bound $a=a_{\rm sing}$ in the physical domain of $a$.
$a=a_{\rm sing}$ is a curvature singularity where $a$ and ${\dot a}$ are finite but ${\ddot a}$ blows up.
In the case where 
\begin{align}
\sigma^2 \le -\frac{2}{\alpha\kappa^4_5}\biggl(1+\frac43 \alpha\Lambda\biggl)^{3/2}
\end{align}
is satisfied, the condition (\ref{recollapse-gb2}) provides the recollapse condition.

\end{document}